\documentstyle[preprint,aps]{revtex}
\begin{document}
\draft
\title{Finite-size anyons and perturbation theory}
\author{Stefan Mashkevich\footnote
{Email: mash@phys.unit.no}}
\address{Institute for Theoretical Physics, 252143 Kiev, Ukraine \\
and \\
Centre for Advanced Study, Drammensveien 78, N-0271 Oslo, Norway}
\preprint{\parbox{5cm}{\begin{flushright} 
Oslo SHS-95-2 \\ hep-th/9511004 \end{flushright}} }
\date{\today}
\maketitle
\begin{abstract}
We address the problem of finite-size anyons, i.e.,
composites of charges and finite radius magnetic flux tubes.
Making perturbative calculations in this problem meets
certain difficulties reminiscent of those in the problem
of pointlike anyons.
We show how to circumvent these difficulties
for anyons of arbitrary spin.
The case of spin $1/2$ is special because it allows
for a direct application of perturbation theory,
while for any other spin, a redefinition of the wave
function is necessary.
We apply the perturbative algorithm to the $N$-body problem,
derive the first-order equation of state and discuss some examples.
\end{abstract}

\section{Introduction}
The topology of two-dimensional
space allows for existence of anyons \cite{Leinaas77,Wilczek82},
particles with statistics intermediate between bosonic
and fermionic. However, the real world being
three-dimensional, no particles with inherent anyonic
statistics can exist; in reality such statistics 
can only arise effectively by means of an interaction.
Remarkably, a composite made of a charge $e$ and a magnetic flux $\Phi$
is an anyon \cite{Wilczek82}, because interchanging two such composites 
multiplies their wave function by $\exp(i\pi\alpha)$,
where $\alpha = e\Phi/2\pi$, in the spirit of the
Aharonov-Bohm effect. Such composites arise naturally in 
Chern-Simons field theory, where, the magnetic field
being proportional to the charge density,
a point charge is at the same time a point flux.
Quasiparticle excitations in the fractional quantum Hall
effect are anyons \cite{Hall} just due to
the charge-flux interaction. 

Since anyons are not fundamental particles but composites or
quasiparticle excitations, they should have finite size.
For example, the size of the FQHE excitations is
of the order of the magnetic length \cite{asw}.
Also, in field theory, if the gauge field
Lagrangian is the Chern-Simons term {\it plus\/} some other
term, a point charge generates a magnetic field smeared over
a finite region. The simplest example is Maxwell-Chern-Simons
theory \cite{Shizuya90,Zinovjev94}. Clearly, if the radius
of the flux tube, which is essentially the inverse photon mass,
is much smaller than any other distance scale
(mean interparticle distance and/or thermal wavelength), 
then the particles are effectively ideal (pointlike) anyons. In the
opposite limiting case, one has particles in a magnetic field
without any change of statistics. Hence there is
``distance dependent statistics'' \cite{Mash93}: If the particles
themselves are, say, bosons, they behave like bosons
when being close together, but like anyons when being far apart.

One more parameter, the size of the anyons, being present
makes quantum mechanical problems more complicated.
The two-body problem was considered
in Refs.~\cite{Zinovjev94,Mash93,Trugen92,Mash93B},
where also some general results and conjectures for the
$N$-body problem can be found.
Since even fewer results can be derived exactly
than for ideal anyons, it is natural to try using perturbation theory
for small values of the statistical parameter. 
For ideal anyons, however, perturbation
theory gives senseless infinite results in the $s$-wave
sector close to bosonic statistics, due to the singular
nature of the anyonic interaction. 
There are several methods to improve the situation,
which are in fact equivalent to each other \cite{Ouvry94}
and may be reduced to the following: Modify
the original Hamiltonian $H_N$ by adding a repulsive contact
interaction, i.e., consider
$H'_N = H_N + c\sum_{j<k}^{1,N}\delta^2({{\bf r}_{jk}})$;
this does not affect the 
exact solutions, but for a special value of $c$
all the divergences cancel, leading to the correct
finite result \cite{McCabe91,Sen,Amelino}.
This value, as was demonstrated in \cite{Comtet95}, corresponds
to assigning to the particles a spin $1/2$,
interacting with the magnetic field inside the flux tube
(cf.~\cite{Hagen}). It is also possible (but not necessary)
to redefine the wave function
in such a way that the terms leading to divergences cancel already
in the Hamiltonian (in addition, also the three-body
interaction terms of $H_N$ then cancel).

Consider the situation in more detail.
Since the nature of the difficulty is connected with two-body
interaction, it is in fact sufficient to restrict oneself to
the two-body problem. In the $L=0$ sector of the relative Hamiltonian 
of two ideal anyons there are two linearly independent solutions,
one regular and the other one singular at $r=0$ ($r$ is the
interparticle distance) but still
normalizable. A generic wave function will behave for small
$r$ as $r^\alpha + \kappa r^{-\alpha}$, where $\kappa$ has to be
fixed as a boundary condition at $r=0$,
which corresponds to choosing one of the possible self-adjoint
extensions of the Hamiltonian \cite{Bourdeau}.
Adding a repulsive contact term excludes the singular solution,
i.e., fixes $\kappa=0$. However, this holds for any value of $c$,
while perturbation theory works for one special value only.
On the other hand, if the singular flux tube is understood as
a limiting case of a finite radius one, then normally only the
regular solution will survive, even with no contact term at all.
In fact, the contact term has to be regularized
itself---and thereby given a physical meaning---and the
result (even in the singular limit) may depend on
the regularization chosen. A simple and natural way is to ascribe
to the particles a magnetic moment, or spin, which couples to the
magnetic field inside the flux tube. The solution will then
exhibit a continuous dependence on the value of the spin, but in
the singular limit the problem becomes scale invariant, and therefore
the dimensionful parameter $\kappa$ can only tend to $\infty$ or to $0$
(unless a distance scale is introduced by hand \cite{Bourdeau},
which will not be considered here). That is, the solution can only
tend to the purely singular one (as we will see, this happens
for spin $+1/2$, where $+$ means attractive, i.e., parallel to
the flux tube) or to the purely regular one (for any other spin).

Our present goal is to reproduce these results within 
(first-order) perturbation theory, in order to make possible
its application to the $N$-body problem.
It works best for spin $\pm1/2$, considered
in detail in Ref.~\cite{Comtet95}, but does not work
directly for any other spin. Analyzing the two-body problem,
we show how the wave function has to be redefined in order that
the first-order result be in agreement with the exact solution.
We then move on to the $N$-body problem. Here again, spin $\pm1/2$
is most simple in the
sense that it is apparently the only case allowing for a cancellation of
the three-body terms; however, our algorithm always works in the sense
that it allows to get rid of the singularities. We apply it
to derive the first-order perturbative equation of state, using
the second-quantized formalism. The limiting cases of small and large
flux tube radius reproduce the ideal anyon and the mean-field
equations of state, respectively, and an example illustrates the
interpolation between the two.

\section{The two-body problem}
Consider the problem of two finite-size anyons.
Let $m$ be their mass, $e$ the charge and $\sigma$ the spin
(more precisely, the projection of the spin on
the $z$ axis, which can be positive or negative),
$\alpha$ the statistics parameter at infinite distance
and $\varepsilon(r)$ the
function describing the flux profile (which we assume to be
radially symmetric), so that
\begin{equation}
\Phi(r) = \frac{2\pi\alpha}{e}\varepsilon(r),
\qquad B(r) = \frac{\alpha}{er} \varepsilon'(r)
\label{1}
\end{equation}
are the flux through the circle of radius $r$ and the
magnetic field at a distance $r$, respectively; 
one has $\varepsilon(\infty) = 1$, and we will assume in the sequel
that $\varepsilon(0) = 0$, i.e., there is no singular flux
at the center (even if the magnetic field still may be singular).
As well as in Ref.~\cite{Comtet95}, spin does not appear from
a relativistic formulation but is introduced by hand.
The magnetic moment, which couples to the magnetic
field, is $\mu=\sigma\frac{e}{m}$ \cite{note}. 
Then the radial part of the relative Hamiltonian, with
a harmonic potential added (the latter, as usually, serves only
for discretizing the spectrum and is irrelevant for the essence
of the matter), is
\begin{equation}
{\cal H} = \frac{1}{m} \left( -\frac{d^2}{dr^2} - \frac{1}{r}\frac{d}{dr}
+ \frac{[L - \alpha\varepsilon(r)]^2}{r^2} 
-\frac{2\sigma\alpha\varepsilon'(r)}{r} \right)
+ \frac{1}{4}m\omega^2r^2;
\label{2}
\end{equation}
$L$ is the angular momentum, and we assume the ``bare'' particles to be
bosons, whence $L$ must be even.
We now wish to treat the $\alpha$ dependent terms as a perturbation.
The difficulties arise for $L=0$,
when the unperturbed wave functions do not vanish at the origin. 
In fact, it is enough to consider the ground state.
For $\alpha=0$, its wave function is
\begin{equation}
\psi_0(r) = \sqrt{m\omega} \exp \left( -\frac{1}{4} m\omega r^2 \right)
\end{equation}
and its energy is $\omega$. Assume, for simplicity,
that $\varepsilon(r)=1$ for $r>R$ (that is, there is no magnetic
field outside the circle of radius $R$), where $R$ is the size of
the anyons. Then the perturbative
correction from the third term of (\ref{2}) is
\begin{eqnarray}
\Delta E &\;&= \left\langle \psi_0 \right|
\frac{\alpha^2\varepsilon^2(r)}{mr^2}
\left| \psi_0 \right\rangle \nonumber \\ &&= 
\int_0^\infty \frac{\alpha^2\varepsilon^2(r)}
{mr^2} m\omega \exp \left(-\frac{1}{2}m\omega^2r^2 \right) r \, dr 
\nonumber \\ &&= \int_0^R {} + \alpha^2 \omega \int_R^\infty
\frac{1}{r} \exp \left(-\frac{1}{2}m\omega^2r^2 \right) \, dr
\nonumber \\ &&
\mbox{\raisebox{-1.5ex}{$\stackrel
{\textstyle \longrightarrow}{\scriptstyle R\to0}$}}
\; \alpha^2\omega (-\ln R
+ {\rm finite\;terms}),
\label{7}
\end{eqnarray}
diverging in the singular limit $R \to 0$. Therefore,
except possibly for some special values of $\sigma$,
the ground state of $\cal H$ for small enough $R$
cannot be reproduced perturbatively in a straightforward way.
(In fact, for $R=0$ the exact result, as we will see, is
$\Delta E = |\alpha| \omega$ for any $\sigma\neq+1/2$.)

\section{Exact solution}
To be able to check the perturbative results, let us first
find the exact ground state energy of ${\cal H}$.
Symmetry with respect to a change $(\alpha,\sigma) \to (-\alpha,-\sigma)$
is present, so it is enough to consider $\alpha\ge0$, which is
what we will assume, unless otherwise specified.
For $r>R$, the wave function with the
correct behavior at infinity is
\begin{equation}
\psi_>(r) = r^\alpha U \left( \frac{1}{2}
\big(1 + \alpha - \frac{E}{\omega} \big), 1+\alpha;
\frac{1}{2} m\omega r^2 \right)
\exp \left(-\frac{1}{4}m\omega r^2 \right),
\label{psig}
\end{equation}
where $U$ is the confluent hypergeometric function.
The energy $E$ can be determined by writing the boundary condition at $r=R$,
\begin{equation}
\frac{\psi'(R)}{\psi(R)} = \frac{\lambda\alpha}{R}
\label{bc}
\end{equation}
and substituting first $\psi_>(r)$ and then $\psi_<(r)$,
the wave function for $r<R$,
thereby obtaining two equations connecting $E$ and $\lambda$.
To make the point clear, let us introduce two restrictions:
(i) $\alpha \ll 1$, since we are interested in perturbation theory;
(ii) $q\equiv m\omega R^2/2 \ll 1$,
since we are now concerned about the singular limit, in which $q\to 0$.
In this approximation\footnote{We will, however, continue to refer
to the result as ``exact'', just to stress that it is obtained
directly from the Schr\"odinger equation.},
the first equation in question reads
\begin{equation}
\frac{E}{\omega}= 1 + \alpha \frac
{1+\lambda - (1-\lambda)q^\alpha}
{1+\lambda + (1-\lambda)q^\alpha}.
\label{beta}
\end{equation}
Note that substituting the asymptotic form
$\psi(r)=r^\alpha+\kappa r^{-\alpha}$ into (\ref{bc}) yields
the relation $\kappa = R^{2\alpha}(1-\lambda)/(1+\lambda)$.
Therefore in the singular limit there are only two possibilities:
If $\lambda\to-1$ (and $1+\lambda$ tends to zero faster than
$R^{2\alpha}$, which will be true for small enough $\alpha$),
then the purely singular solution survives and $E\to(1-\alpha)\omega$,
in any other case the regular solution survives and 
$E\to(1+\alpha)\omega$. On the other hand,
for perturbation theory it is the factor $q^\alpha$ in
(\ref{beta}) that is troublesome, because its perturbative expansion
$q^\alpha = 1 + \alpha\ln q + \frac{\alpha^2\ln^2 q}{2} + \cdots$
needs more and more terms as $q$ tends to $0$.
This factor cancels out for $\lambda\to\pm1$, and
one expects just these two situations to cause no difficulties for
perturbation theory.

To proceed with the exact solution, one has to supply a function
$\varepsilon(r)$ for $r<R$. The simplest choice here is
$\varepsilon(r) = r^2/R^2$, corresponding to  
the magnetic field being uniform
inside the circle of radius $R$ \cite{Trugen92,Mash93B,Comtet95}.
Then (\ref{2}) becomes
\begin{equation}
{\cal H}_< = \frac{1}{m} \left( -\frac{d^2}{dr^2} 
- \frac{1}{r}\frac{d}{dr} \right)
+ \frac{1}{4}m\tilde\omega^2r^2 + E_0,
\label{hless}
\end{equation}
where
\begin{equation}
\tilde\omega^2 = \omega^2 + \frac{\alpha^2}{4m^2R^4}, \qquad
E_0 = - \frac{4\sigma\alpha}{mR^2}.
\label{ome}
\end{equation}
The corresponding solution---regular at the origin, since there is
no singular flux tube---is
\begin{equation}
\psi_<(r) = {}_1\!F_1 \left( \frac{1}{2}
\big(1 - \frac{E-E_0}{\tilde\omega} \big), 1;
\frac{1}{2} m\tilde\omega r^2 \right)
\exp \left(-\frac{1}{4}m\tilde\omega r^2 \right),
\label{psiless}
\end{equation}
and substitution into (\ref{bc}) yields
\begin{equation}
\lambda = \frac{1}{\alpha} 
\frac{qE/\omega + 2\sigma\alpha}{qE/2\omega - 1}.
\label{lam}
\end{equation}
For small enough $q$ one may put $\lambda=-2\sigma$,
because $q$ tends to zero faster than $q^\alpha$. 
Consequently, the singular solution arises for $\sigma=+1/2$
(i.e., attractive), the regular one does for any other $\sigma$.
Perturbation theory experiences difficulties for any $\sigma\neq\pm1/2$.
In particular, for the spinless case $\sigma=0$
one gets $E/\omega = 1 + \alpha(1-q^\alpha)/(1+q^\alpha)$, i.e.,
\begin{equation}
\Delta E =
- \alpha \omega \tanh \left( \frac{\alpha}{2}\ln q \right).
\label{tanh}
\end{equation}
For $q\to0$, this tends to $\Delta E = \alpha\omega$,
but to see this, one has to take into account {\it all}\/
orders in $\alpha$.

\section{Perturbation theory}
Let us now come back to perturbation theory.
Following the idea of \cite{McCabe91,Sen},
in order to get rid of the singularities we redefine the
wave function as
\begin{equation}
\psi(r) = f(r) \tilde{\psi}(r),
\label{ansatz}
\end{equation}
where $f(r)$ is to be fixed at our convenience. Then the Hamiltonian
acting on $\tilde\psi(r)$ is
\begin{equation}
\tilde{\cal H} = \tilde{\cal H}_0 + \tilde{\cal H}_1
+ \tilde{\cal H}_2,
\label{9}
\end{equation}
where
\begin{eqnarray}
\tilde{\cal H}_0 & = & \frac{1}{m} 
\left( -\frac{d^2}{dr^2} - \frac{1}{r}\frac{d}{dr} \right)
+ \frac{1}{4}m\omega^2r^2,
\label{10} \\ 
\tilde{\cal H}_1 & = & -\frac{2}{m} \frac{f'(r)}{f(r)}
\frac{d}{dr},
\label{11} \\ 
\tilde{\cal H}_2 & = & \frac{1}{m} \left( -\frac{f''(r)}{f(r)}
- \frac{1}{r} \frac{f'(r)}{f(r)} + \frac{\alpha^2\varepsilon^2(r)}{r^2}
- \frac{2\sigma\alpha\varepsilon'(r)}{r} \right).
\label{12} 
\end{eqnarray}
The purpose of the redefinition is to have the dangerous $1/r^2$
term canceled, by an appropriate choice of $f(r)$.
Asymptotically for $r\to\infty$, when $\varepsilon(r)=1$, demanding
it to be canceled by any linear combination of the first two terms of
$\tilde{\cal H}_2$ will make $f(r)$ be a power function,
and then each of these two terms will itself be
proportional to $1/r^2$ and therefore should be canceled as well.
As we will see, the correct way is to demand that all
four terms be canceled, i.e., that $\tilde{\cal H}_2\equiv0$, or
\begin{equation}
r^2 f''(r) + rf'(r) - \big[ \alpha^2 \varepsilon^2(r) 
- 2\sigma\alpha r\varepsilon'(r) \big] f(r) = 0.
\label{difeq}
\end{equation}
This is the essence of the perturbative algorithm for our problem:
Find $f(r)$ from (\ref{difeq}), choosing
the solution which is nonsingular at the origin,
and then regard $\tilde{\cal H}_1$ as a perturbation.
This algorithm will yield the correct result in the singular limit.

Remarkably, for $\sigma=\mp1/2$ there is
a universal solution to (\ref{difeq}),
\begin{equation}
f(r) = \exp \left[ \pm \alpha\int_0^r \frac{\varepsilon(r')}{r'}
\: dr' \right],
\label{anss}
\end{equation}
which is precisely the ansatz considered in \cite{Comtet95}.
Then one obtains ${\cal H}_1 = \mp \frac{2}{m} \frac{\alpha
\varepsilon(r)}{r} \frac{d}{dr}$, which yields the correct
first-order result; in the singular limit it is
$\Delta E = \pm\alpha\omega$. In fact, here this redefinition is
not necessary; acting directly with the original problem
(\ref{2}) is more complicated but still possible, because
the divergent first-order contribution from the $1/r^2$ term
turns out to be canceled by the
second-order contribution from the spin term,
while the first-order contribution from the latter gives the
correct answer. 

For arbitrary spin, there is apparently no universal solution.
However, use can be made of the fact that $\alpha\ll1$, to
construct an approximate one. Coming back to the uniform
magnetic field model, one has
\begin{equation}
\left\{ \begin{array}{ll}
r^2 f''(r) + rf'(r) - [\frac{\alpha^2r^4}{R^4}
-\frac{4\sigma\alpha r^2}{R^2}]f(r) = 0, & \qquad r<R, \\
r^2 f''(r) + rf'(r) - \alpha^2 f(r) = 0, & \qquad r>R.
\end{array} \right.
\label{syst}
\end{equation}
It is easy to see
that $f_<(r) = 1-\sigma\alpha\frac{r^2}{R^2}$ is the nonsingular
solution to the first equation, neglecting terms of the order
$\alpha^2$. The general solution to the second one is
$f_>(r)=C_1r^\alpha+C_2r^{-\alpha}$, and matching $f$ and $df/dr$
at $r=R$ yields
\begin{equation}
f(r) = \left\{ \begin{array}{ll}
1 - \sigma\alpha\frac{r^2}{R^2}, & \qquad r<R, \\
\frac{1-\sigma(2+\alpha)}{2}\left(\frac{r}{R}\right)^\alpha +
\frac{1+\sigma(2-\alpha)}{2}\left(\frac{R}{r}\right)^\alpha,
& \qquad r>R.
\end{array} \right.
\label{sol}
\end{equation}
Now the first-order correction from ${\cal H}_1$ is
\begin{eqnarray}
\Delta E & = & \left\langle \psi_0 \right| \tilde{\cal H}_1
\left| \psi_0 \right\rangle \nonumber \\
& = &2\sigma\alpha\omega\frac{qe^{-q}+e^{-q}-1}{q}
+ \alpha\omega\int_q^\infty
\frac{(1-2\sigma)x^\alpha - (1+2\sigma)q^\alpha}
{(1-2\sigma)x^\alpha + (1+2\sigma)q^\alpha}e^{-x}\:dx.
\label{defi}
\end{eqnarray}
For $q\ll1$, the first term can be neglected, and in the second
term one may replace $x^\alpha$ by 1 (this is legal unless
$|\ln x| \gtrsim 1/\alpha$, but the main contribution to the
integral is given by the values $x\sim1$), and then the lower
limit of integration can be replaced by $0$, which yields the
exact result (\ref{beta}).

It is worthwhile to note that the ``straightforward'' result (\ref{7}),
for the spinless case,
is correct in a sense; indeed, if in (\ref{tanh}) one fixes $q$
and goes to the limit $\alpha \to0$, one does get (\ref{7}). 
This is natural: There being no singularity, the straightforward
perturbation theory works for small enough $\alpha$; but its
range of validity shrinks to zero in the limit $q\to0$, and to
get the expression which remains valid in this limit, one has
to redefine the wave function as described above.

Concerning the excited states, for the ones with $L=0$ and the
radial quantum number $n\neq0$ one has to apply the same
algorithm, because its sense is to take care of the short-distance
behavior of the wave function, which is independent
of $n$. For $L\neq0$, perturbation theory is directly applicable,
because already the unperturbed wave functions vanish at the
origin, although making the redefinition will do no harm.
Thus, the algorithm can be applied for all the states, which is
essential for the second-quantized formalism.

\section{The $N$-body problem and the perturbative equation of state}
The main goal of perturbation theory is its application to the
$N$-body problem. The Hamiltonian in our case reads
\begin{equation}
H_N = \frac{1}{2m} \sum_{j=1}^N
\left[ \left( -i\frac{\partial}{\partial {\bf r}_j}
- e\sum_{k\neq j}{\bf A}({\bf r}_{jk}) \right)^2
- 2\sigma e\sum_{k\neq j}B(r_{jk}) \right] + V,
\label{Hn}
\end{equation}
where
\begin{equation}
{\bf A}({\bf r}) = \frac{\alpha}{e}
\frac{{\bf e}_z \times{\bf r}}{r^2}\varepsilon(r)
\end{equation}
is the vector potential and $B(r)$ is the magnetic field as in
(\ref{1}); ${\bf r}_{jk} = {\bf r}_j - {\bf r}_k$, and
$r_{jk} = |{\bf r}_{jk}|$. The wave function now is to be
transformed as
\begin{equation}
\psi({\bf r}_1, \ldots,{\bf r}_N) = 
\left( \prod_{j<k}^{1,N} f(r_{jk}) \right)
\tilde\psi({\bf r}_1, \ldots,{\bf r}_N)
\end{equation}
with $f(r)$ determined as above. Then the Hamiltonian acting
on $\tilde\psi$ is
\begin{eqnarray}
\tilde{H}_N &=\displaystyle
\frac{1}{2m} \sum_{j=1}^N & \left[ 
-\frac{\partial^2}{\partial{\bf r}_j^2}
+2i\alpha\sum_{\stackrel{\scriptstyle k\neq j}{\phantom{k\neq l}}}
 \frac{{\bf e}_z \times{\bf r}_{jk}}{r_{jk}}
 \varepsilon(r_{jk})\frac{\partial}{\partial{\bf r}_j}
-2\sum_{\stackrel{\scriptstyle k\neq j}{\phantom{k\neq l}}}
\frac{f'(r_{jk})}{f(r_{jk})}
 \frac{{\bf r}_{jk}}{r_{jk}}\frac{\partial}{\partial{\bf r}_j}\right.
\nonumber \\
&& \:\:\left.+\sum_{\stackrel{\scriptstyle k,l\neq j}{k\neq l}}\left(
 -\frac{f'(r_{jk})f'(r_{jl})}{f(r_{jk})f(r_{jl})}
 + \alpha^2\frac{\varepsilon(r_{jk})\varepsilon(r_{jl})}{r_{jk}r_{jl}}
 \right) \frac{{\bf r}_{jk}{\bf r}_{jl}}{r_{jk}r_{jl}}\right] + V,
\end{eqnarray}
where Eq.~(\ref{difeq}) has been taken into account.
This contains two-body as well as three-body interaction terms.
If $\sigma=\pm1/2$ then, because of Eq.~(\ref{anss}), the three-body
terms cancel, otherwise they do not. This certainly
makes the problem more complicated, however these terms are
of second order in $\alpha$ and they do not produce any
singularities. Therefore in the first order one omits them
and considers the second and the third terms
as perturbation. The first-order contribution of the second term,
in fact, vanishes, and by virtue of the symmetry of the wave
function the contribution of the third term can be represented as
\begin{equation}
\Delta E = \frac{N(N-1)}{2} \langle \psi_{\rm sym} |
\tilde{\cal H}_1({\bf r}_1, {\bf r}_2)
| \psi_{\rm sym} \rangle,
\end{equation}
where
\begin{equation}
\tilde{\cal H}_1({\bf r}_1, {\bf r}_2) = -\frac{2}{m}
\frac{f'(r_{12})}{f(r_{12})}
\frac{{\bf r}_{12}}{r_{12}}\frac{\partial}{\partial{\bf r}_1}
\end{equation}
[cf.~(\ref{11})].

It is now possible to derive the (first-order) perturbative
equation of state for finite-size anyons near both Bose and
Fermi statistics. (In the second case, as well as for bosons
outside the $s$-wave sector, the redefinition of the
wave function is not necessary but does not change the result.)
The simplest way is to use the second quantized formalism.
The starting point is
the expression for the first-order correction to
the thermodynamic potential \cite{dVO92}
\begin{eqnarray}
\Omega_1 & = & \int_0^\beta d\beta_1 \int d^2{\bf r}_1 \, d^2{\bf r}_2
\,\tilde{\cal H}_1 ({\bf r}_1, {\bf r}_2) \left[
\left\{ \psi^\dagger({\bf r}_1,\beta_1) \psi({\bf r}_1,\beta_1) \right\}
\left\{ \psi^\dagger({\bf r}_2,\beta_1) \psi({\bf r}_2,\beta_1) \right\}
\right. \nonumber \\ & & \pm \left.
\left\{ \psi^\dagger({\bf r}_1,\beta_1) \psi({\bf r}_2,\beta_1) \right\}
\left\{ \psi^\dagger({\bf r}_2,\beta_1) \psi({\bf r}_1,\beta_1) \right\}
\right],
\label{eos1}
\end{eqnarray}
where 
$\left\{ \psi^\dagger({\bf r}_1,\beta_1) \psi({\bf r}_2,\beta_2) \right\}$
is the one-particle thermal Green function, and the upper/lower sign refers
to bosons/fermions.
Taking into account the explicit form of $\tilde{\cal H}_1$
(note that the derivative acts on $\psi^\dagger$ only),
it is possible to do the spatial integration by parts,
and the surface term will vanish.
There is an expression 
(see \cite{dVO92} for details)
\begin{equation}
\left\{ \psi^\dagger({\bf r}_1,\beta_1) \psi({\bf r}_2,\beta_1) \right\}
= -\sum_{s=1}^\infty (\pm z)^s G({\bf r}_1,{\bf r}_2; s\beta),
\label{eos2}
\end{equation}
where $z$ is the fugacity and
\begin{equation}
G({\bf r}_1,{\bf r}_2; \beta) = \frac{1}{\lambda^2}
\exp \left[ -\frac{\pi ({\bf r}_1-{\bf r}_2)^2}{\lambda^2} \right]
\label{eos3}
\end{equation}
is the one-particle plane wave thermal Green function
in the thermodynamic limit,
$\lambda = \sqrt{2\pi\beta/m}$ being the thermal wavelength.
Equation (\ref{eos1}) then turns into
\begin{equation}
\Omega_1 = \alpha \frac{V}{2\lambda^2} 
\sum_{n=1}^\infty c_n (\pm z)^n
\label{eos4}
\end{equation}
with
\begin{eqnarray}
c_n & = & \frac{1}{2} \sum_{s=1}^{n-1} \frac{I_{s,n-s}}{s(n-s)}, \\
I_{st} & = & \frac{1}{\alpha} \int_0^\infty 
dr \, \frac{d}{dr} \left[ r\frac{f'(r)}{f(r)} \right]
\left[ 1 \pm \exp\left( -\frac{s+t}{st} \frac{\pi r^2}{\lambda^2} \right)
\right] . 
\label{eos5}
\end{eqnarray}
According to Eq.~(\ref{difeq}), one has
\begin{equation}
\begin{array}{l}
f(r) \mathop{\rightarrow}\limits_{r\to0} 1, \\
f(r) \mathop{\rightarrow}\limits_{r\to\infty} C_1r^\alpha + C_2r^{-\alpha},
\end{array} \label{asymp}
\end{equation}
therefore the integral above is
in fact proportional to $\alpha$, and $c_n \propto 1$.
Now, one has for the pressure
\begin{eqnarray}
P\beta & = & -(\Omega_0 + \Omega_1)/V \nonumber \\
& = & \frac{1}{\lambda^2} \sum_{n=1}^\infty \left( \pm\frac{1}{n^2}
- \alpha c_n \right) (\pm z)^n
\label{eos7}
\end{eqnarray}
[$\Omega_0 = \mp (V/\lambda^2) \sum_{n=1}^\infty (\pm z)^n/n^2$
being the unperturbed thermodynamic potential] and for the density
\begin{eqnarray}
\rho & = & z\frac{\partial}{\partial z} (P\beta) \\  
& = & \frac{1}{\lambda^2}\left[ \mp \ln(1\mp z) - \alpha \sum_{n=1}^\infty
n c_n (\pm z)^n \right];
\end{eqnarray}
neglecting $\alpha^2$, the solution for $z$ is
\begin{equation}
z = z_0 + \alpha(1\mp z_0) \sum_{n=1}^\infty n c_n (\pm z_0)^n,
\label{eos11}
\end{equation}
where
\begin{equation}
z_0 = \pm[1 - \exp(\mp \lambda^2\rho)]
\end{equation}
is the unperturbed fugacity. Equations (\ref{eos7}) and (\ref{eos11})
explicitly give $P$ as a function of $\rho$, i.e., the equation of state,
which is thus is obtained for any flux profile in terms of integrals.

Introducing the virial expansion,
\begin{equation}
P\beta = \rho[1 + A_2(\rho\lambda^2) + A_3(\rho\lambda^2)^2 + \ldots] \, ,
\end{equation}
the first few virial coefficients are

\parbox{13cm}{\begin{eqnarray*}
A_2 & = & \mp\frac{1}{4} + \alpha (-\frac{1}{2}c_1 + c_2), \\
A_3 & = & \frac{1}{36} \pm \alpha (\frac{1}{3}c_1 - 2c_2 + 2c_3), \\
A_4 & = & \alpha (-\frac{1}{8}c_1 + \frac{7}{4}c_2 - \frac{9}{2}c_3
 + 3c_4), \\
A_5 & = & -\frac{1}{3600} \pm \alpha( \frac{1}{30}c_1 - c_2 + 5c_3 - 8c_4
 + 4c_5).
\end{eqnarray*}} \hfill 
\parbox{1cm}{\begin{eqnarray} \label{eos14} \end{eqnarray}}

\section{Examples}
Consider some particular cases. We will now drop the
condition $\alpha > 0$. According to (\ref{asymp}), the main term
of $f(r)$ at $r\to\infty$ is proportional to $r^{|\alpha|}$ unless
the corresponding $C$ vanishes, whence it is proportional to
$r^{-|\alpha|}$ [in particular, it is so when
$\sigma\frac{\alpha}{|\alpha|} = +1/2$ and $\varepsilon(r)
=\eta(r)$]. In the first case,
it is easy to see from (\ref{eos5}) that
\begin{equation}
I_{st} = \left\{ \begin{array}{ll} (1 \pm 1) \frac{\alpha}{|\alpha|} \qquad &
 \mbox{for} \quad R \ll\lambda, \\ 
\frac{\alpha}{|\alpha|} \qquad & \mbox{for} \quad R \gg\lambda.
\end{array} \right.
\end{equation}
This leads to simple results:
\begin{eqnarray}
R \ll\lambda : & \qquad & A_2 = \mp1/4 + \frac{1\pm1}{2}|\alpha|, \\
R \gg\lambda : & \qquad & A_2 = \mp1/4 + \frac{1}{2}|\alpha|,
\end{eqnarray}
and the higher virial coefficients are unaffected. The first equation
is the well-known first-order result for ideal anyons \cite{dVO92},
the second one is a mean-field result \cite{mf}---indeed, in that
limit the magnetic field gets smeared over the whole volume. 
In the second case---that is, essentially for attractive spin $1/2$---there
is an additional minus sign in the correction terms, so that the
second virial coefficient can become lower than the bosonic one
(cf.~\cite{Blum}).

The transition between the two limiting regimes can be observed
on a simple model where
an explicit expression for $c_n$ is within reach.
Let $\sigma = -1/2$ and $\varepsilon(r) = 1-\exp(-r^2/R^2)$,
so that the magnetic field $B(r)=(2\alpha/eR^2) \exp(-r^2/R^2)$
is Gaussian.
Then, by virtue of (\ref{anss}), $rf'(r)/f(r) = \varepsilon(r)$,
and one gets
\begin{equation}
I_{st} = 1 \pm \frac{1}{1+ \xi^2 \frac{s+t}{st}}, \\
\end{equation}
\begin{eqnarray}
c_n  = \frac{\gamma + \psi(n)}{n} \pm \frac{1}{2\sqrt{n^2 + 4n\xi^2}}
& & \left[ \psi \left( \frac{n + \sqrt{n^2 + 4n\xi^2}}{2} \right)
       - \psi \left( \frac{2 - n + \sqrt{n^2 + 4n\xi^2}}{2} \right) \right.
\nonumber \\ & & \left. {}
       - \psi \left( \frac{n - \sqrt{n^2 + 4n\xi^2}}{2} \right)
       + \psi \left( \frac{2 - n - \sqrt{n^2 + 4n\xi^2}}{2} \right) \right],
\end{eqnarray}
where
\begin{equation}
\xi = \sqrt{\pi} \frac{R}{\lambda},
\end{equation}
$\gamma$ is Euler's constant, $\psi(x)=\Gamma'(x)/\Gamma(x)$.
[The same expressions will of course be valid for any other $\sigma$
and $\varepsilon(r)$ that lead to the same $f(r)$, through Eq.~(\ref{difeq}).]

At the lowest order in $\xi$ and $1/\xi$, respectively, the virial
coefficients are

\parbox{13cm}{\begin{eqnarray*}
A_2 & = & \mp\frac{1}{4} + \alpha(\frac{1\pm1}{2} \mp \xi^2), \\
A_3 & = & \frac{1}{36} + \alpha\frac{\xi^2}{2}, \\
A_4 & = & \mp \alpha\frac{\xi^2}{12}, \\
A_5 & = & -\frac{1}{3600}, \\ \\
A_2 & = & \mp\frac{1}{4} + \alpha(\frac{1}{2} \pm \frac{1}{4\xi^2}), \\
A_3 & = & \frac{1}{36} + \alpha\frac{1}{6\xi^2}, \\
A_4 & = & \pm \alpha\frac{1}{16\xi^2}, \\
A_5 & = & -\frac{1}{3600} + \alpha\frac{1}{60\xi^2}.
\end{eqnarray*}} \hfill 
\parbox{1cm}{\begin{eqnarray} 
 \label{ll1} 
 \\ \nonumber \\ \nonumber \\ \nonumber \\ \nonumber \\
\label{gg1}\end{eqnarray}} \\
The Bose and Fermi $\xi$ dependent terms are equal to each other for
odd coefficients and have opposite signs for even ones, and for $\xi\ll1$
they in fact vanish for all odd coefficients but the third.
Plots of the first five virial coefficients as functions of $\xi$ near
Bose statistics, showing the intermediate behavior, are displayed in Fig.~1.

Note that there is no symmetry with respect to $\alpha\to-\alpha$ here;
indeed, the spin is repulsive for $\alpha>0$ but attractive for $\alpha<0$.

\section{Discussion}
Let us summarize the main points of our reasoning.
In the problem of finite-size anyons that was considered,
there are no singularities and therefore all results, exact
or perturbative, are finite.
Therefore, had it been possible to calculate the perturbative
corrections to all orders (and provided the series converged),
the exact result for the problem at hand could be obtained
without any special treatment.
What is achieved by the redefinition of the wave function---in
a way that actually takes into account the short-distance behavior
of the exact solution---is the result being obtained at lower order
of perturbation theory than it would be without this redefinition.
For spin $1/2$, it is obtained at first order instead of
second; for any other spin, at first order instead of infinite.

There is a certain subtlety concerning the
transformation of the Hamiltonian.
The correct singular limit is $\varepsilon(r) = \eta(r)$ (the
step function) rather than $\varepsilon(r) = 1$ \cite{Comtet95},
so that the spin term in (\ref{2}) becomes a contact term
$-\frac{4\pi\sigma\alpha}{m}\delta^2({\bf r})$.
For $\sigma=\mp1/2$, the function $f(r)$
defined by (\ref{anss}) tends in the singular limit
to $f_0(r) = r^{\pm\alpha}$, and it is the equality
$\Delta\ln f_0(r) = \pm2\pi\alpha\delta^2({\bf r})$
that ensures the cancellation
of the contact term when the Hamiltonian is redefined
\cite{Ouvry94,Comtet95}. Now, for arbitrary $\sigma$,
the function $f(r)$ defined by (\ref{sol}) also tends to $r^\alpha$,
but only {\it pointwise\/}; an elementary calculation
shows that $\Delta\ln f(r) \to -4\pi\sigma\alpha\delta^2({\bf r})$,
thereby ensuring the correct transformation of the Hamiltonian.
Again, spin $1/2$ is singled out, in the sense that it is in this case
only that the limiting transition under the operator $\Delta$
is legal.

The encoding of the short-distance behavior of the two-body
wave function in perturbation theory allows for a perturbative
treatment of the $N$-body problem. Again, this turns out to
be most simple for spin $1/2$ (because the three-body terms
cancel), but still possible, in principle, for arbitrary spin.
The perturbative equation of state is obtained in an explicit
form and shows smooth interpolation between the two limiting
cases, the ideal anyon one and the mean filed one.

\bigskip
Numerous stimulating discussions
with St\'ephane Ouvry are gratefully acknowledged.
I am also thankful to Giovanni Amelino-Camelia for useful remarks.
Parts of this work were done at the theory division of the
Institut de Physique Nucl\'eaire, Orsay
and at the Centre for Advanced Study at the Norwegian Academy of
Science and Letters, Oslo,
thanks to which are due for kind hospitality and support.

\newpage
{\large \bf Figure caption} \\[1cm]

Fig.~1a--d. The virial coefficients $a_2$ through $a_5$
as functions of $\xi$ in the Gaussian model, for $\alpha = 0.1$.


\begin{references}
\bibitem{Leinaas77} J.M.~Leinaas, J.~Myrheim,
  Nuovo Cimento B {\bf 37}, 1 (1977).
\bibitem{Wilczek82} F.~Wilczek,
  Phys.~Rev.~Lett. {\bf 48}, 1144 (1982), {\it ibid.}~{\bf 49}, 957 (1982).
\bibitem{Hall} R.B.~Laughlin, Phys.~Rev.~Lett. {\bf 50}, 1953 (1983);
  B.~Halperin, {\it ibid.}~{\bf 52}, 1583 (1984).
\bibitem{asw} D.~Arovas, J.R.~Schrieffer, F.~Wiclzek,
  Phys.~Rev.~Lett. {\bf 53}, 722 (1984).
\bibitem{Shizuya90} K.~Shizuya, H.~Tamura,
  Phys.~Lett. B {\bf 252}, 412 (1990).
\bibitem{Zinovjev94} G.~Zinovjev, S.~Mashkevich, H.~Sato, JETP {\bf 78},
  105 (1994) [Russian original: Zh.~Eksp.~Teor.~Fiz.~{\bf 105}, 198 (1994)].
\bibitem{Mash93} S.~Mashkevich, Phys.~Rev. D {\bf 48}, 5953 (1993).
\bibitem{Trugen92} C.A.~Trugenberger, Phys.~Lett. B {\bf 288}, 121 (1992).
\bibitem{Mash93B} S.~Mashkevich, G.~Zinovjev,
  JETP {\bf 82}, 813 (1996).
  [Russian original: Zh.~Eksp.~Teor.~Fiz. {\bf 109}, 1512 (1996).]
\bibitem{Ouvry94} S.~Ouvry, Phys.~Rev. D {\bf 50}, 5296 (1994).
\bibitem{McCabe91} J.~McCabe, S.~Ouvry, Phys.~Lett. B {\bf 260}, 113 (1991);
  A.~Comtet, J.~McCabe, S.~Ouvry, Phys.~Lett. B {\bf 260}, 372 (1991).
\bibitem{Sen} D.~Sen, Nucl.~Phys. B {\bf 360}, 397 (1991).
\bibitem{Amelino} G.~Amelino-Camelia, Phys.~Rev. D {\bf 51}, 2000 (1995).
\bibitem{Comtet95} A.~Comtet, S.~Mashkevich, S.~Ouvry,
  Phys.~Rev. D {\bf 52}, 2594 (1995).  
\bibitem{Hagen} C.R.~Hagen, Phys.~Rev. D {\bf 52}, 2466 (1995).
\bibitem{Bourdeau} C.~Manuel, R.~Tarrach,
  Phys.~Lett. B {\bf 268}, 222 (1991); M.~Bourdeau, R.D.~Sorkin,
  Phys.~Rev. D {\bf 45}, 687 (1992); G.~Amelino-Camelia, hep-th/9502105.
\bibitem{note} In fact, we are tacitly assuming that the gyromagnetic
  ratio for the spin is $g=2$ (while $2\sigma$ might in principle
  be fractional) \cite{gyro}.
  More generally, $\sigma$ in our formulas would of course be replaced
  by $g\sigma/2$, with the conclusions changing correspondingly if $g\neq2$.
  In particular, for relativistic {\it fundamental}\/ particles
  (not composites) of integer or half-integer spin $\sigma\neq0$
  there is the result $g=1/\sigma$ \cite{Hagen70}, so that in this case 
  $\sigma = \pm1,\pm3/2,\pm2,\ldots$ are all equivalent to 
  $\sigma = \pm1/2$.
  I thank C.R.~Hagen for pointing this out.
\bibitem{gyro} C.~Chou, V.P.~Nair, A.P.~Polychronakos,
  Phys.~Lett. B {\bf 304}, 105 (1993);
  G.~Gat, R.~Ray, Phys.~Lett. B {\bf 340}, 162 (1994).
\bibitem{Hagen70} C.R.~Hagen, W.J.~Hurley,
  Phys.~Rev.~Lett. {\bf 24}, 1381 (1970).
\bibitem{dVO92} A.~Dasni\`eres de Veigy, S.~Ouvry, Phys.~Lett. B {\bf 291},
  130 (1992).
\bibitem{mf} S.~Viefers, Cand.~Sci.~thesis (Oslo, 1993).
\bibitem{Blum} T.~Blum, C.R.~Hagen, S.~Ramaswamy, Phys.~Rev.~Lett. {\bf 64},
  709 (1990).
\end{references}
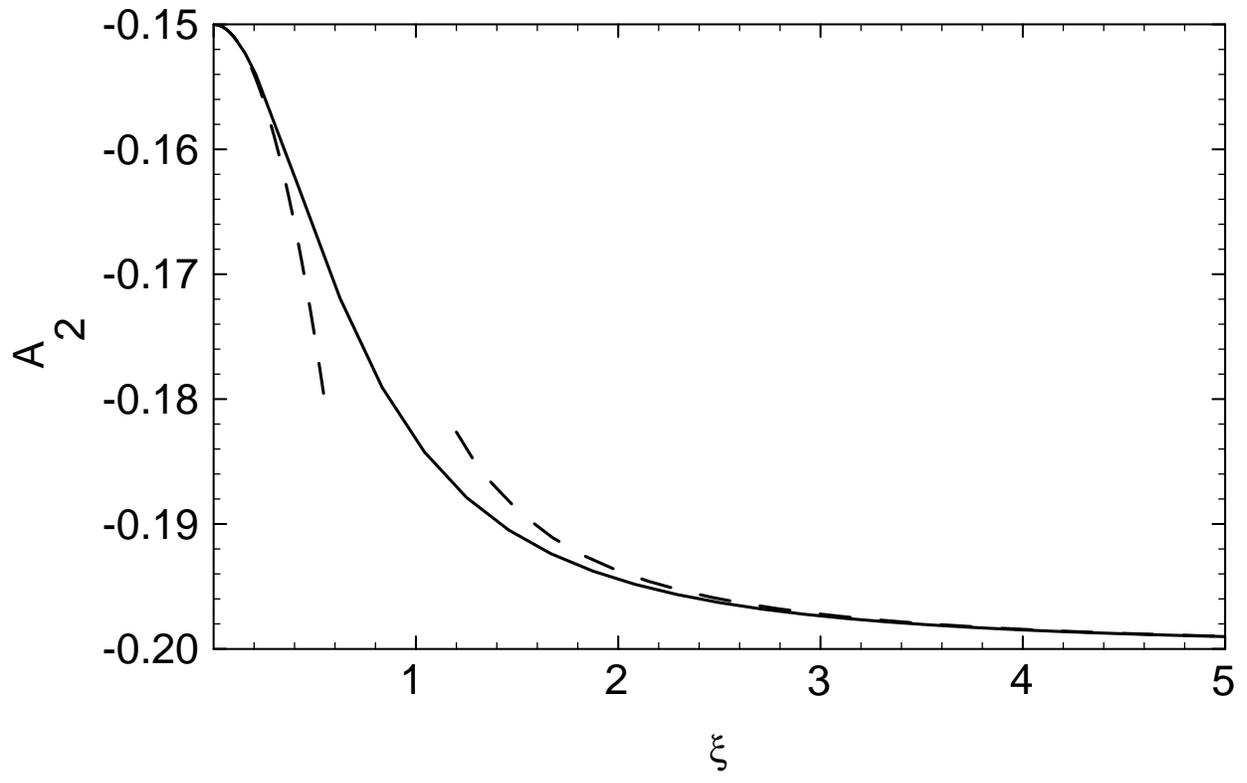
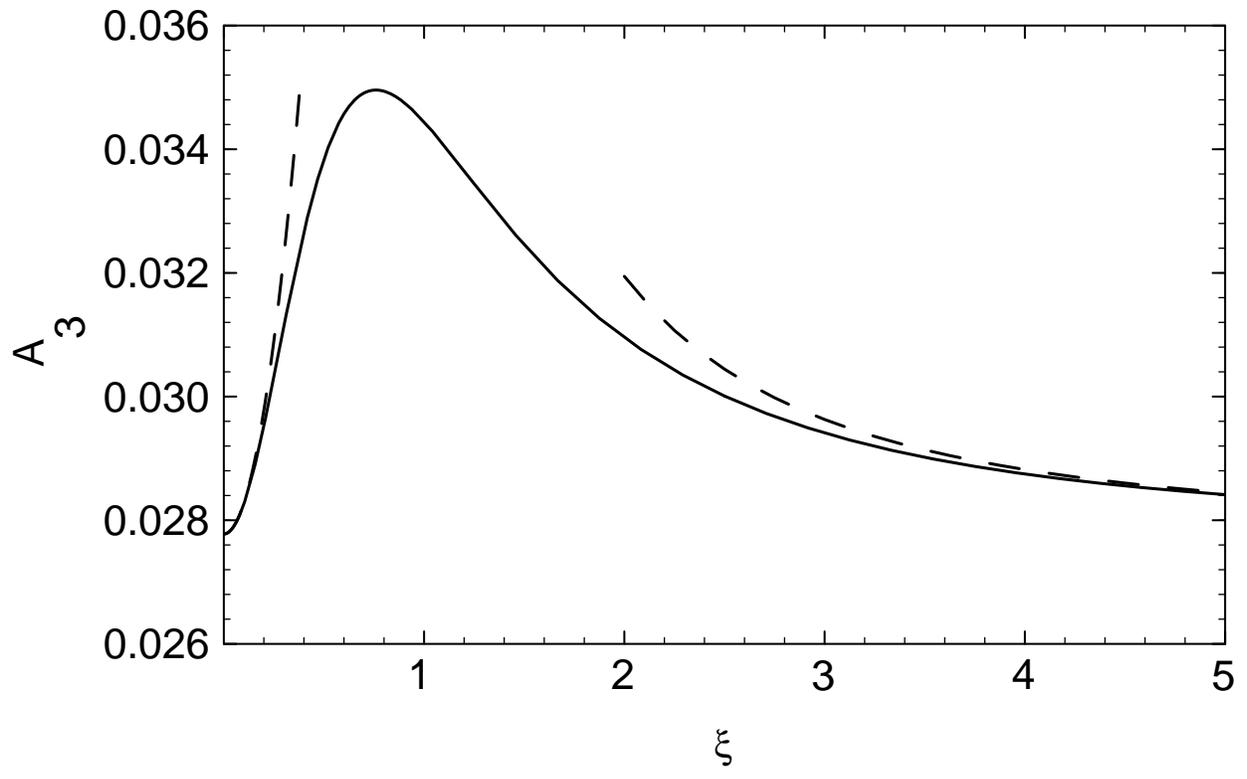
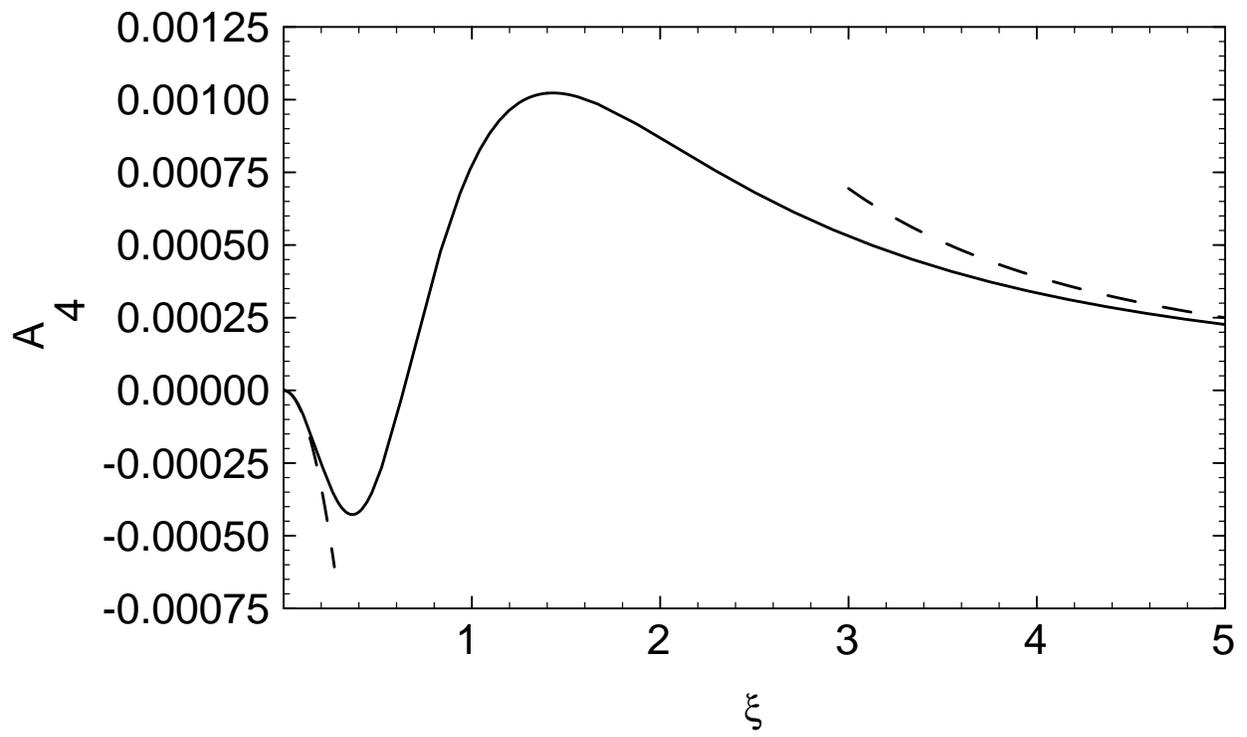
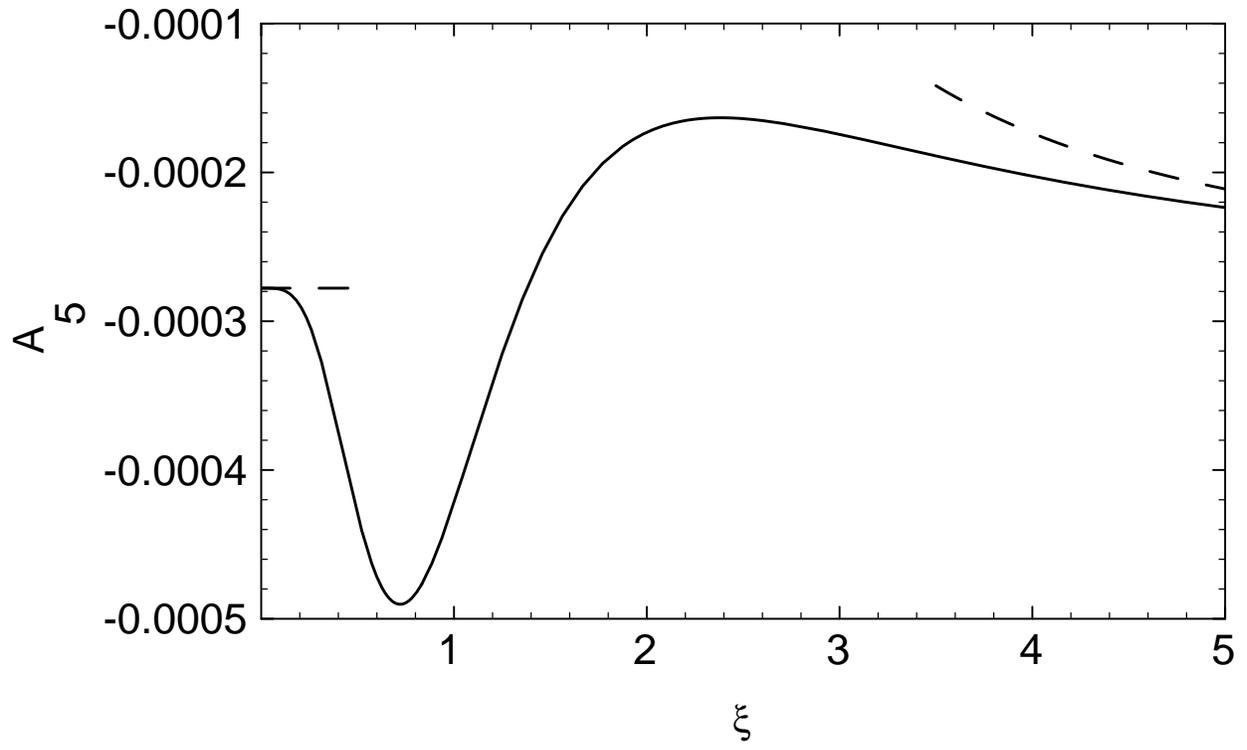
\end{document}